\documentclass{aa}
\usepackage{graphicx}
 
\def\msun{\hbox{$M_\odot$}}
\begin{document} 
\title{HST observations of the metal rich globular clusters NGC\,6496 and
NGC\,6352\thanks{Based on observations with the NASA/ESA Hubble Space
Telescope, obtained at the Space Telescope Science Institute, which is
operated by AURA for NASA under contract NAS5-26555}}

\author{Luigi Pulone \inst{1}, Guido De Marchi \inst{2}, Stefano
Covino\inst{3} and Francesco Paresce \inst{4}}

\offprints{L. Pulone}

\institute{
INAF, Osservatorio Astronomico di Roma, Via di Frascati 33, I--00040
Monte Porzio Catone (RM), Italy
\and
European Space Agency, Space Telescope Operations Division, 3700 San
Martin Drive, Baltimore MD 21218, USA
\and
INAF, Osservatorio Astronomico di Brera, Via Bianchi 46, I--23807
Merate (LC), Italy
\and 
European Southern Observatory, Karl-Schwarzschild-Strasse 2, D--85748,
Garching, Germany}

\date{Received ; Accepted }
\titlerunning{HST observation of NGC\,6496 and NGC\,6352}
\authorrunning{L. Pulone et al.}

\abstract{ 
Deep exposures of the metal-rich globular clusters NGC\,6496\ and
NGC\,6352 were obtained with the WFPC2 camera on board the Hubble Space
Telescope (HST) through the F606W and F814W filters. The resulting
colour-magnitude diagrams (CMD) reach down to absolute magnitude
$M_{814} \simeq 10-10.5$, approximately 5 magnitudes below the main
sequence (MS) turn-off (TO). The MS of the two clusters are sharp and
well defined and their fiducial lines overlap almost exactly throughout
this range. Their colour is, however, more than $0.1$\,mag redder than
the MS fiducial line of the prototype metal-rich globular cluster
47\,Tuc (NGC\,104), after proper correction for the relative distances
and reddening. This provides solid empirical evidence of a higher metal
content, which is not surprising if these objects belong indeed to the
bulge as their present location suggests.  A good fit to the upper part
of the MS of both clusters is obtained with a $10$\,Gyr-old theoretical
isochrone from Baraffe et al.  (1998) for a metallicity of
[M/H]=$-0.5$, but at lower luminosities all models depart considerably
from the observations, probably because of a deficiency in the
treatment of the TiO opacity.  The luminosity functions (LF) obtained
from the observed CMD are rather similar to one another and show a peak
at $M_{814}\simeq 9$. The present day mass functions (PDMF) of both
clusters are derived down to $M_{814}\simeq 10.5$ or $m \simeq
0.2$\,M$_\odot$ and are consistent with power-law indices $\alpha=0.7$
for NGC\,6496 and $\alpha=0.6$ for NGC\,6352. The PDMF of NGC\,104 is
twice as steep in the same mass range ($\alpha=1.4$). We investigate
the origin of this discrepancy and show that it can be understood if
the two clusters contain a considerably higher fraction of primordial
binaries amongst their MS population, similar to that expected in the
bulge.  We briefly discuss the implications of this finding on the
process of star and binary formation and on the universality of the
IMF.
  
}
\maketitle

\section{Introduction}

Since the work of Zinn \& West (1984), Armandroff \& Zinn (1988) and
Armandroff (1989), the disc population of the Galactic globular
clusters (GC) has been recognised as a different subsystem, both
dynamically and chemically, compared with that of the halo clusters. A
detailed knowledge of these subsystems is important for our
understanding of the fundamental properties of GC such as the
concentration parameter, the shape of the mass function (MF), their age
and chemical composition, since they correlate with the Galactocentric
distance (Meylan \& Heggie 1997) and/or height above the Galactic
plane.  Studying the disc GC, although difficult because of the
foreground star contamination and high extinction due to their low
Galactic latitude, could shed light on the disruption mechanisms by
dynamical friction and bulge shocks, which deeply modify their MF in
time. Furthermore, investigating the physical differences between their
stellar population and that of the halo clusters could improve our
understanding of the formation process of our Galaxy.  In this paper,
we present the analysis of the deep CMD of NGC\,6496 and NGC\,6352, two
clusters in the direction of the bulge and putative members of this
subgroup, based on observations collected with the WFPC2 on board the
HST.

NGC\,6496 was originally believed to be a member of the disc system of
GC, but Richtler et al. (1994) questioned this classification. They
suggested instead that NGC\,6496, together with two other clusters,
NGC\,6624 and NGC\,6637, could be halo clusters with strongly inclined
orbits. NGC\,6496 lies in the Southern sky at
$\alpha_{2000}=17^{o}59\arcmin02\arcsec$ and
$\delta_{2000}=-44^{0}15\arcmin54\arcsec$. Its Galactic coordinates are
$l=348.02^{o}$ and $b=-10.1^{o}$. A distance of $R_G=4.3$\,kpc from the
Galactic centre and $Z_G=-2.0$\,kpc below the Galactic plane (Harris
1996), place this object at the edge of the bulge.  Notwithstanding the
fact that the surrounding field shows no strong stellar foreground
variation, its low Galactic latitude has hampered the observation of
this cluster from the ground. Indeed, most of the contradictory
estimates as to its metallicity, reddening, and Galactocentric distance
probably arise mainly from the difficulty of observing objects through
the dusty and crowded Galactic disc.

Armandroff (1988) presented the first CMD of NGC\,6496 in which the
photometry reached $\sim 2$ magnitudes below the horizontal branch
(HB), disclosing for the first time the usual red arm of the metal-rich
clusters. The extinction towards NGC\,6496 is uncertain, with estimates
ranging between $E(B-V)\,=\,0.09$ and $E(B-V) = 0.24$ (Armandroff 1988;
Zinn \& West 1984; Burstein \& Heiles 1982; Richtler et al. 1994).

The metal content of NGC\,6496 has also been controversial for a long
time. Friel \& Geisler (1991), in their work on the disc GC based on
Washington photometry, used the reddening value of $0.19$ by Armandroff
(1988) and derived for NGC\,6496 a metallicity of [Fe/H]$=-1.05$.
Richtler et al. (1994) measured its metallicity on the basis of the
morphological characteristics of the CMD, obtaining [Fe/H]$=-1.0$.
Their value is in agreement with the metallicity coming from Washington
photometry, but substantially lower than that given by Zinn \& West
(1984), i. e. [Fe/H]$ = -0.48$, and adopted by Armandroff (1988). The
discrepancies in the metallicity scale amongst different authors might
be explained by the intimate connection between reddening and
metallicity estimates.

NGC\,6352 is located at $\alpha_{2000}=17^{o}25\arcmin29\arcsec$ and
$\delta_{2000}=-48^{o}25\arcmin22\arcsec$. Its Galactic coordinates
(${\it l}=341.4^{o}$, ${\it b}=-7.2^{o}$) and distance from the
Galactic centre and plane ($R_G=3.3$\,kpc, $Z_G=-0.7$\,kpc; Harris
1996) place this objects at the edge or within the bulge. Based on this
information and on radial velocity and metallicity measurements,
NGC\,6352 has been recognised, since Zinn (1985) and Armandroff (1989),
as an unambiguous member of the disc group of globular clusters.
Sarajedini \& Norris (1994) presented CCD ground-based photometry for
NGC\,6352 reaching $\sim 2$ magnitudes below the level of its HB,
disclosing the red clumped morphology typical of metal-rich clusters.
By simultaneous estimate of both metallicity and reddening, using
theoretical stellar evolutionary tracks, they obtain $[Fe/H]=-0.51$, in
the Zinn \& West metallicity scale, and $E(B-V)=0.24$.

Fullton et al. (1995) observed NGC\,6352 with the WFPC1 on board HST,
obtaining a CMD reaching $\sim 3$ magnitudes below the MS TO. Their
main results are that NGC\,6352 is slightly more metal rich than
47\,Tuc and that it has the same age as 47\,Tuc within the errors of
their photometry. These results imply that the primordial gas might
have collapsed in a disc configuration at an early epoch of the
formation of our Galaxy.

In light of all the uncertainties affecting the determination of the
extinction, distance and metallicity of these clusters, the skeptical
reader might wonder as to how meaningful and practically useful it is
to classify these objects as being members of one or another subgroup
of the GC. We believe that the availability of new, powerful
instruments such as the HST and VLT should call for a critical
re-evaluation of the semi empirical methods on which such
classifications were based.  In the following sections we provide solid
observational data in the form of high precision CMD of NGC\,6496 and
NGC\,6352, covering almost the full range of stellar masses from the TO
to $\sim 0.2 \msun$. We compare them with the CMD of the archetype,
well-studied metal-rich globular cluster 47\,Tuc (NGC\,104) and address
the issue of their metallicity in a way so far never attempted, yet
robust in light of its simplicity.  Furthermore, from the LF we derive
the best fitting underlying PDMF to test the hypothesis as to whether
the IMF could be metal dependent and, therefore, differ here from that
of the metal-poor halo clusters studied by Paresce \& De Marchi
(2000).
 
\section{Observations and data analysis}

All the exposures of both clusters NGC\,6496 and NGC\,6352, as
summarised in Table\,1, have been taken in fine lock mode, with the
WFPC2 camera on board the HST, through the F606W and F814W
broad-band filters.

\begin{table}
\centering 
\caption{HST observations of the clusters NGC\,6496 and NGC\,6352}
\begin{tabular}{cccrc} \hline
Cluster & Date UT & Filter & Exp. time (s) & Nr. exp. \\
\hline      
NGC\,6496 & 1999 Apr 1 & F606W & 30  &  1\\ 
     &      & F606W & 40 &  1\\ 
     &      & F606W & 1100 & 2 \\ 
     &      & F606W & 1300 & 4 \\ 
     &      & F814W & 30 &  1\\
     &      & F814W & 40  &  1\\
     &      & F814W & 1100 & 2 \\ 
     &      & F814W & 1300 & 4 \\ \hline
NGC\,6352 & 1998 Mar 3 & F606W & 30  & 3 \\ 
     &      & F606W & 40 &  1\\  
     &      & F606W & 1300 & 8 \\  
     &      & F814W & 30 & 3 \\
     &      & F814W & 40  &  1\\
     &      & F814W & 1200 & 2 \\ 
     &      & F814W & 1300 & 6 \\ \hline	 
\end{tabular}
\vspace{0.5cm}
\label{tab1}
\end{table}

The images of NGC\,6496 were obtained by targetting a region located
$\sim 1.3\arcmin$ East of the cluster centre, at coordinates
$\alpha_{2000}=17^{o}59\arcmin12\arcsec$ and
$\delta_{2000}=-44^{o}15\arcmin59\arcsec$. The field imaged in our
programme is close to the half-light radius ($r_h=1\farcm9$; Harris
1996), where the local MF is expected to be less affected by mass
segregation resulting from internal dynamical evolution (Richer et al.
1991).

\begin{figure}
\centering
\resizebox{\hsize}{!}{\includegraphics{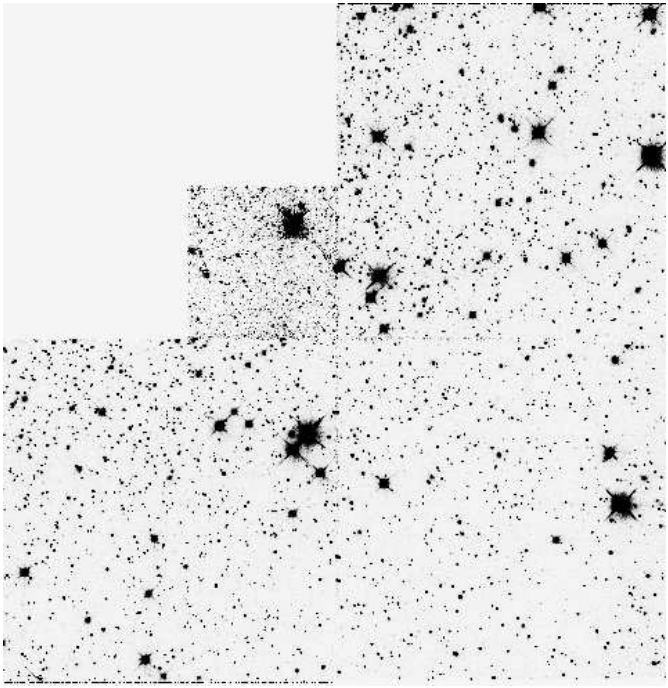}}
\caption{Negative image of the cluster NGC\,6496 of 40\,sec duration in
the F814W band}
\label{fig1}
\end{figure}

\begin{figure}
\centering
\resizebox{\hsize}{!}{\includegraphics{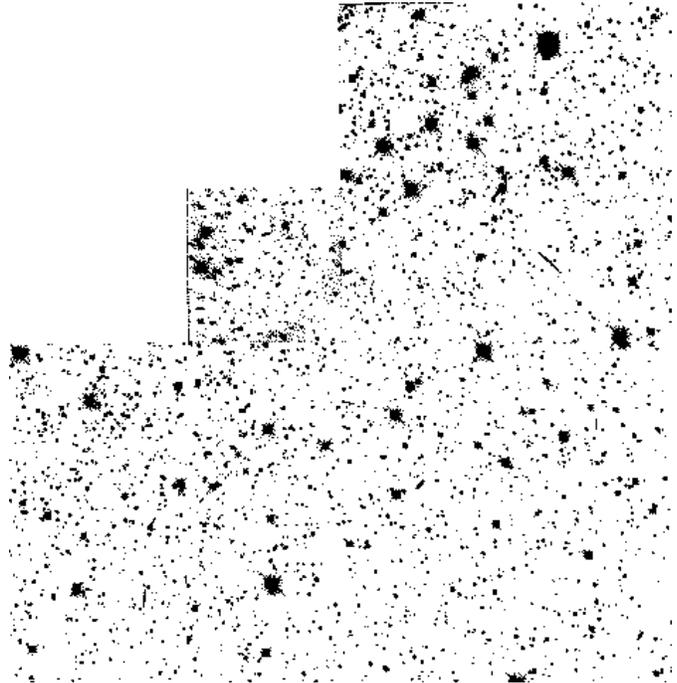}}
\caption{Negative image of the cluster NGC\,6352 of 40\,sec duration 
in the F814W band} 
\label{fig2}
\end{figure}

The field observed in NGC\,6352 is located at
$\alpha_{2000}=17^{o}25\arcmin29.2\arcsec$ and
$\delta_{2000}=-48^{o}25\arcmin20.2\arcsec$, or about $ 2\arcmin$ North
of the cluster centre, at a distance corresponding to the half-light
radius ($r_h=2\farcm0$; Harris 1996). The mosaic images of the fields
observed in NGC\,6496 and NGC\,6352 are shown in Figures \ref{fig1} and
\ref{fig2}, respectively, and are based on two $40$\,sec F814W
exposures (see Table\,1).

The raw images were processed using the standard HST pipeline
calibration.  The long exposures taken through the same filter were
combined to remove cosmic ray hits and to improve the signal-to-noise
ratio of the data. The total equivalent exposure time of the combined
images of NGC\,6496 corresponds to $7400\,s$ in each band. As regards
NGC\,6352, the total exposure time equals $10400\,s$ in the F606W
band, and $10200\,s$ in F814W. The same procedure has been applied to
the short exposures, thus producing two $70\,s$ images in both bands
for NGC\,6496 and two $130\,s$ images in both bands for NGC\,6352.

Due to the large number of pixel affected by cosmic rays, satellite
tracks and detector defects that affected the raw frames of NGC\,6352,
however, we decided to derive the final scientific frames for these
observations by taking their median rather than the usual mean. As is
well known (see for instance the DAOPHOT users' manual; Stetson 1987),
the median is very effective in removing outliers, but from a
statistical point of view the median of $n$ frames is roughly
comparable to the mean of only $2/3 $ of the frames. Therefore, as far
as the star detection algorithm is concerned, our effective exposure
time was reduced to $6930\,s$ and $6800\,s$ in F606W and F814W,
respectively, compared to the ones obtained simply by summing all
available frames. The ensuing reduction in the SNR is more than
compensated by the higher accuracy achieved in the photometry, resulting
in a cleaner CMD and tighter MS.

Photometry was carried out using the standard IRAF DAOPHOT package.
The automated star detection routine {\it daophot.daofind} was applied
to the averaged frames, by setting the detection threshold at $5\sigma$
above the local background level. We have then visually examined all
the detected objects in this way. False identifications, due
essentially to the diffraction spikes of a number of saturated stars,
were removed from the output list. In this way, concerning the long
exposures data set, in the NGC\,6496 field a total of
$650$,\,$2070$,\,$1500$,\,$2300$ objects were detected, respectively in
the $PC$, $WF2$, $WF3$, and $WF4$ chips. As regards NGC\,6352, the
detected objects were $379$,\,$1443$,\,$1279$, and \,$1216$.

The photometric routines {\it phot} and {\it allstar} were used to
measure the fluxes on the combined frames. The PSF was modelled by
averaging moderately bright stars. We calculated the aperture
corrections for the four chips, by using a diameter of $0.5\arcsec$ and
translated the final instrumental magnitudes into the VEGAMAG
photometric system by adopting the updated zero points listed in the
March 1998 edition of the {\it HST Data Handbook}. It should be noted
that, thanks to the low level of crowding in these images, aperture
photometry would have provided equally robust results.

\section{The colour--magnitude diagram}

\begin{figure*}
\centering
\includegraphics[width=16cm]{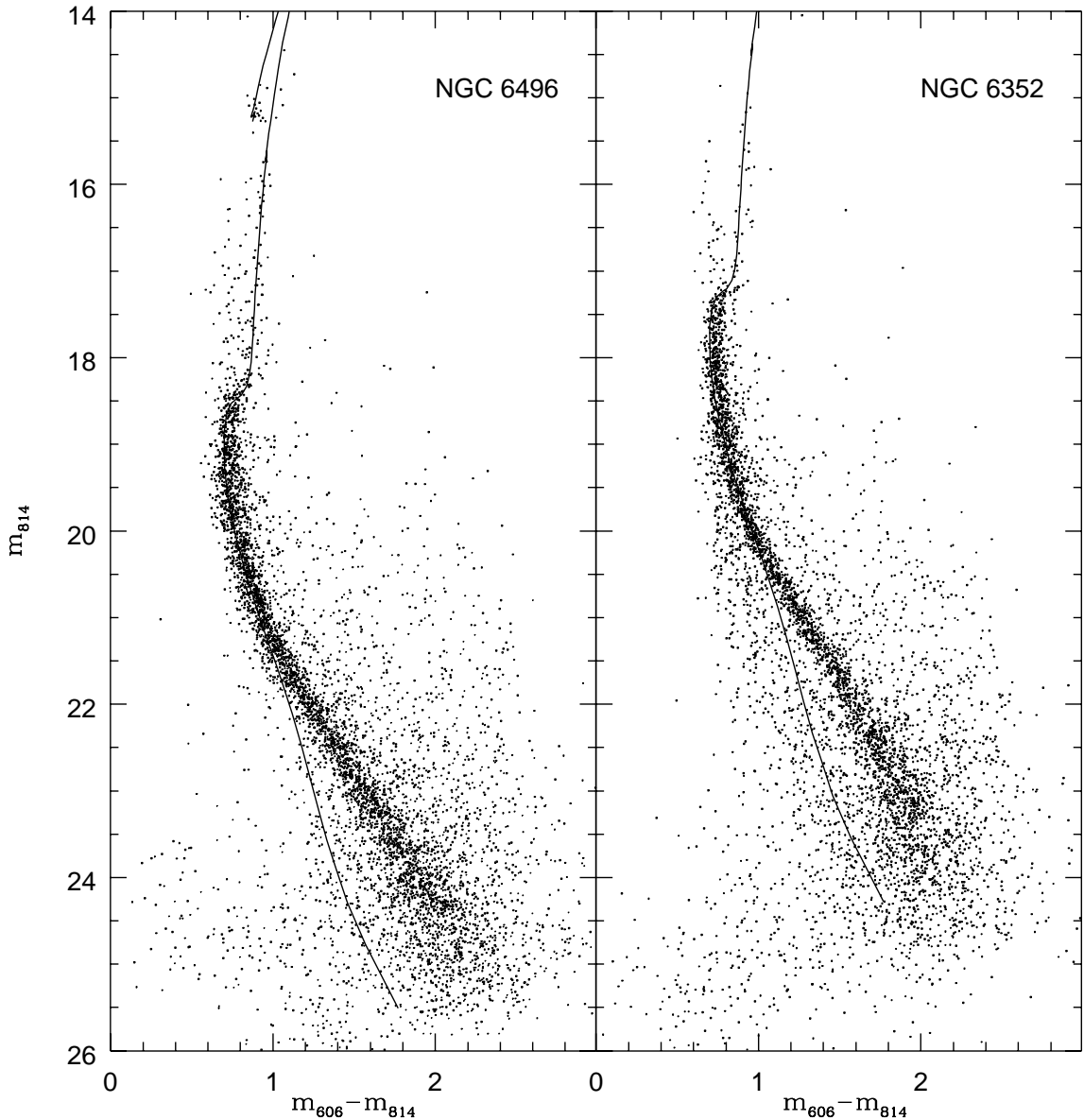}
\caption{CMD of NGC\,6496 (left panel)
and NGC\,6352 (right panel). Each CMD contains all the detected objects
in the short and long exposures. A 14 Gyr isochrone  by Girardi et al.
(2002) has been overlapped as explained in the text}
\label{fig3}
\end{figure*}

Figure\,\ref{fig3} shows the $m_{814}$, $m_{606}-m_{814}$ CMD of
NGC\,6496 and NGC\,6352, for all the objects measured in the combined
short and long exposures. The photometry reaches $m_{814} \sim 25.5$
and the MS of the clusters are sharp, well defined and stand out
clearly from the background contamination down to $m_{814} \sim 24$.
 
In the long exposures, saturation occurs at $m_{814} < 19.5$ in the WF
chips and at about one magnitude brighter for the PC (thanks to the
higher plate-scale of this camera). Above these brightness limits, the
data in Figure\,\ref{fig3} come therefore from the short exposures. The
MS changes slope at $m_{814} \simeq 23$ for NGC\,6496 and $m_{814}
\simeq 22$ for NGC\,6352, where it becomes slightly steeper. The change
of slope of the MS is due to the effect of dissociation of $H_2$
molecules in the envelope of the star, and to the consequent lowering
of the adiabatic gradient (Copeland et al. 1970). This has already been
extensively discussed by D'Antona (1995) and D'Antona \& Mazzitelli
(1996).

Although the F606W and F814W filters are not the ideal choice for
estimating the age of the clusters, the area around their TO is well
reproduced by a 14 Gyr old isochrone by Girardi et al. (2002) for a
metallicity of $Z=0.008$ and Helium content $Y=0.25$ (see
Figure\,\ref{fig3}). This fit suggests a distance modulus
$(m-M)_{o}=14.8 \pm 0.1$ for NGC\,6496 and $(m-M)_{o}=14.6 \pm 0.1$ for
NGC\,6352, respectively, with colour excess $E(B-V)=0.25$ for both
clusters. The distance modulus of NGC\,6496 is consistent with that of
Richtler et al. (1994), whereas that of NGC\,6352 complies with the
value published by Harris (1996) for this cluster. The extinction in
the F606W and F814W bands has been approximated following the law of
Cardelli et al. (1989) and the uncertainties around the colour excess
towards the two clusters is discussed in detail in Section\,4. The CMD
of NGC\,6496 reveals a short, stubby horizontal branch (HB) at
$m_{814}\simeq 15.5$, a characteristic which is typical of metal rich
clusters and is well reproduced by the theoretical HB track of a $0.69
M_{\odot}$ star as shown by the isochrone.

To determine the MS fiducial points, we have calculated the mean
$m_{606}-m_{814}$ colour for each 0.5 mag bin along the MS, clipping
all the objects lying farther than $\pm2.5\,\sigma$ from the mean
colour value, and then recomputed the mean colour until the convergence
was reached and a fiducial MS mean ridge line was finally defined.
Table\,2 and 3 list, respectively for NGC\,6496 and NGC\,6352, the
average MS colour and the associated standard deviation ($\sigma $) as
a function of the $m_{814}$ magnitude, together with the corrected
stellar counts per bin, and the completeness fraction. The faint end of
the MS of NGC\,6352 is about one magnitude brighter than that of
NGC\,6496, due to the shorter equivalent exposure time of the NGC\,6352
median averaged frames.
 
The MS TO of NGC\,6496 lies at $m_{814} \sim 19.25$. Based on the
transformations linking Johnson's photometry in the $B$ and $V$ bands
to the HST photometric system (Holtzman et al. 1995), one should expect
the TO to be located at $V\simeq 20.15$, i.e. the same value found by
Richtler et al. (1994) in their ground-based photometry of NGC\,6496.
For NGC\,6352 one finds, after correction for reddening and the
transformation to the Johnson's system, that the TO is at $V\simeq
18.8$, in excellent agreement with the value by Fullton et al. (1995).

\section {Metallicity and reddening of NGC\,6496 and NGC\,6352}

Published values of the metallicity of NGC\,6496 cover a wide range
from $[Fe/H]=-0.48$ (Zinn \& West 1984) to $[Fe/H]=-1.05$ (Friel \&
Geisler 1991). This is, however, hardly surprising, since, as discussed
by Richtler et al. (1994), the metallicity estimated by CMD
morphological indicators is usually significantly lower than that
obtained through integrated aperture photometry. (Note, however, that
whilst true at the time when it was made, this statement does not apply
to the comparison of the MS fiducial lines that we discuss below and
which is far more reliable than any integrated property.) On the other
hand, metallicity and reddening estimates are strictly connected.  If
we consider an increase of reddening from $E(B-V)=0.19$ to
$E(B-V)=0.23$, the metallicity evaluated by Friel \& Geisler (1991)
would increase to $[Fe/H]=-0.8$, in better agreement with the estimates
based on CMD features which depend on metallicity, such as the slope
$S$ of the red giant branch (Hartwick 1968). If estimated through the
IR maps of Schlegel et al. (1998), the reddening of NGC\,6496 is
$E(B-V)=0.24$.

The metallicity and reddening estimates of NGC\,6352, according to all
the available measurements based on different methods (Zinn \& West
1984; Gratton 1987; Geisler at al. 1991; Da Costa \& Armandroff 1991;
Fullton et al. 1995; Carretta \& Gratton 1997) indicate that it is
closer in Iron abundance to NGC\,6496 than to 47\,Tuc.  
Following the metallicity scale of Carretta \&
Gratton (1997), the Iron abundance of NGC\,6352 is $[Fe/H]=-0.64$,
whereas for 47\,Tuc they give $[Fe/H]=-0.7$. All the reddening
estimates are in the range $E(B-V) =[0.21, 0.24]$, except for the one
of Schlegel et al. (1998) which, in the direction of NGC\,6352, yields
$E(B-V) = 0.34$.

\begin{figure}
\resizebox{\hsize}{!}{\includegraphics{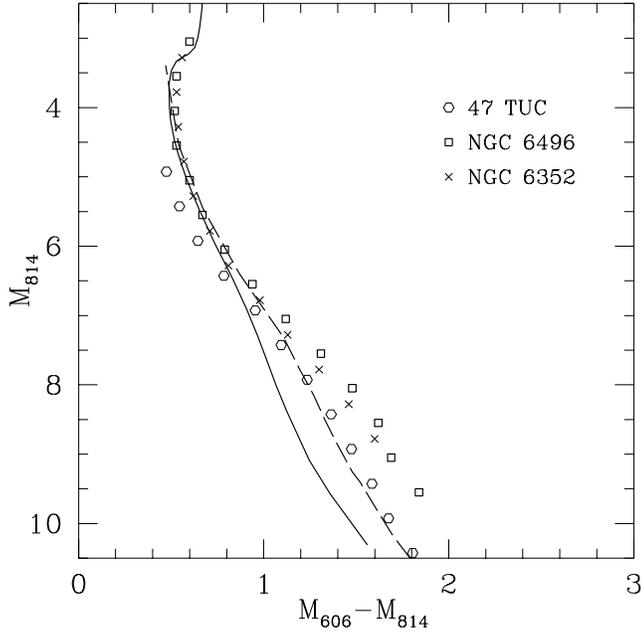}}
\caption{Fiducial MS lines of NGC\,6496 (squares) compared with
that of NGC\,6352 (crosses) and 47\,Tuc (circles). The 14\,Gyr
isochrone of Girardi et al. (2002) for $Z=0.008$ is shown (solid line)
together with the 10\,Gyr isochrone by Baraffe et al.  (1998) for
[M/H]$ =-0.5$ (dashed line)}
\label{fig4}
\end{figure}

To learn more about the physical properties of the stars in these
clusters, we compared the CMD of NGC\,6496 and NGC\,6352 with that of a
reference cluster, i.e. one with known reddening, metallicity, distance
and age, such as NGC\,104 (47\,Tuc). Further information can be derived
from the features of the CMD which reflect specific stellar properties
and their evolution and which can be compared with theoretical
isochrones of various chemical composition and ages as calibrated
through field sub-dwarfs. Figure\,\ref{fig4} shows the fiducial MS lines
of NGC\,6496 (squares) and NGC\,6352 (crosses) as compared
with that of 47\,Tuc (circles), whose photometry comes from
De Marchi \& Paresce (1995b). The dashed line represents the 10\,Gyr
isochrone from the models of Baraffe et al. (1998) for metal content
[M/H]$=-0.5$ in the VEGAMAG photometric system (Isabelle Baraffe 1998,
private communication). For sake of comparison, we also show the
14\,Gyr isochrone of Girardi et al. (2002) for metallicity $Z=0.008$
and $Y=0.25$. The distance modulus and extinction for NGC\,6496 and
NGC\,6352 are those given above (respectively $(m-M)_{o}=14.8$ and
$E(B-V)=0.25$, $13.6$ and $0.25$), whereas for 47\,Tuc we have adopted
$(m-M)_{o}=13.25$ and $E(B-V)=0.04$ from Percival et al. (2002).
There is overall agreement between the models and the observed
sequences for absolute magnitude brighter than $M_{814} \simeq 6.5$.
The deviation of the Baraffe et al. (1998) models close to the TO is
probably due to the younger age of the adopted isochrone, whilst the
discrepancies between the two sets of models and between these and the
data below $M_{814} \simeq 7$ most likely arise because of the
uncertainties in the treatment of the opacity in the atmospheres of
these high metallicity cool stars.  Nevertheless, the isochrone of
Baraffe et al. (1998) better approximates the observed loci of
NGC\,6496 and NGC\,6352. We shall, therefore, adopt the
corresponding M/L relation to derive a MF from the observed LF. 
 
Interestingly, since the MS of NGC\,6496 and NGC\,6352 lie in the same
place of the diagram and have an almost identical shape, one must
conclude (see e.g. Cassisi et al. 2000), that these clusters have the
same chemical composition. Indeed, it is well known that the slope of
the very low mass end of the MS depends strongly on the metallicity
of the stars.  More importantly, however, the comparison in
Figure\,\ref{fig4} is insensitive to errors in the distance and/or
reddening, in that none of them could alter the shape of the MS but
only shift it solidly in the colour--magnitude plane. On this basis,
one is forced to exclude the high reddening value of $E(B-V) = 0.34$,
suggested by the IR dust maps of Schlegel et. al.  (1998), as this
would imply that the metallicity of NGC\,6352 is considerably higher
than that of NGC\,6496, at variance with the overall agreement shown in
Figure\,\ref{fig4}.  However, the large value of $E(B-V)$ measured by
Schlegel et al. (1998) is probably due to dust behind the cluster or
some diffuse infrared background in the direction of NGC\,6352.

\begin{figure}
\resizebox{\hsize}{!}{\includegraphics{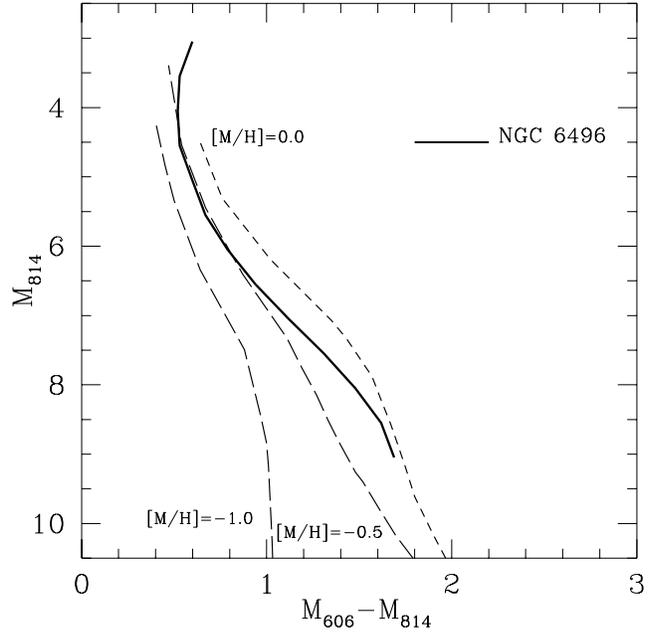}} 
\caption{Theoretical isochrones from Baraffe et al. (1998) (dashed
lines) compared with the MS fiducial line of NGC\,6496 (solid line)}
\label{fig5}
\end{figure}

The fiducial line of 47\,Tuc (circles) is reasonably well fitted by the
[M/H]$=-0.5$ isochrone by Baraffe et al. (1998) and is at least
$0.1$\,mag bluer than those of the two other clusters at bright
magnitudes (the difference grows larger still at lower masses) thus
witnessing its lower metal content.  This conclusion is further
strengthened by the different slope of the MS of 47\,Tuc down to
$M_{814} \simeq 6$.  Such an empirical estimate can be checked against
the predictions of the most recent theoretical models, as we do in
Figure\,\ref{fig5}, where we compare the MF of NGC\,6496 (thick solid
line) and the isochrones of Baraffe et al. (1998) for a stellar
population of 10\,Gyr of age.  The dashed lines represent, from left to
right, three 10\,Gyr isochrones of metal content, respectively,
[M/H]$=-1$, $-0.5$ and and $0.0$. Only the latter (short dashed line)
takes the updated TiO opacities into account (see below).

Firstly, we note that the [M/H]$=-1.0$ isochrone dose not fit at all the
position nor the shape of the NGC\,6496 fiducial line, so the hypothesis
of an intermediate metal content for this cluster (Richtler et al. 1994)
is inconsistent with the present observational and theoretical results.
It is, therefore, extremely unlikely that NGC\,6496 is a halo object.

The [M/H]$=-0.5$ isochrone, which corresponds to [Fe/H]$\simeq -0.7$
when taking into account the overabundance of $\alpha$-elements (Ryan
\& Norris 1991), fits the upper part of the MS rather well, yet it
fails to reproduce the lower portion. The deviation of the models from
the observed stellar locus at the bottom of the MS is a well known
limitation of the theory and most probably stems from the lack of a
proper treatment of the TiO molecular opacity, as extensively discussed
in Baraffe et al. (1998; see also Chabrier 2001). Although a proper
correction is currently not available, we show in the next section that
the adoption of the theoretical [M/H]$=-0.5$ mass-luminosity (M--L)
relation will still provide, in the F814W band alone, reliable
results.  Nevertheless, Figure\,\ref{fig5} allows us to conclude that
the true metal content of NGC\,6352 and NGC\,6496 --- i.e. [M/H], not
just [Fe/H] --- must be at least $0.2 - 0.3$ dex higher than that of
47\,Tuc.

\section{The luminosity and mass functions}

Tables\,2 and 3 list, for each magnitude bin, the completeness
fraction, the corrected LF and the corresponding rms errors coming from
the Poisson statistics of the counting process (all values have been
rounded off to the nearest integer).  In order to derive the LF, and
thus infer the PDMF of the clusters, we must correct the observed
counts for the incompleteness of the sample due to crowding and to
saturated stars, whose bright halo can mask possible faint objects in
their vicinity. Completeness corrections have been estimated by running
artificial star tests in both bands and in the combined images for each
chip. For each $0.5$ magnitude bin we have carried out 10 trials by
adding a fraction of 10\,\% of the total number of objects in that bin
(see Section\,2), and using the PSF from the co-added frames.  These
trials were followed by running the tasks {\it daophot.daofind} and
{\it daophot.allstar}, with the same parameters used in the reduction
of the scientific images to assess the fraction of the objects
recovered by the procedure and the associated photometric errors.

\begin{table} 
\centering
\caption{Main sequence fiducial points, luminosity function and 
completeness for NGC\,6496}
\begin{tabular}{cccccc}
\hline {\em $m_{814}$} & {\em $m_{606}-m_{814}$} & {\em $\sigma$} &{\em
$N_{o}$} & {\em Compl. (\%)}\\ \hline
 18.25 & 0.81 & 0.06 &  80 & 100 \\
 18.75 & 0.74 & 0.03 & 140 &  99 \\
 19.25 & 0.73 & 0.04 & 222 &  99 \\
 19.75 & 0.77 & 0.04 & 248 &  98 \\
 20.25 & 0.81 & 0.05 & 288 &  97 \\
 20.75 & 0.88 & 0.05 & 320 &  96 \\
 21.25 & 1.00 & 0.06 & 325 &  95 \\
 21.75 & 1.15 & 0.09 & 330 &  94 \\
 22.25 & 1.33 & 0.09 & 356 &  91 \\
 22.75 & 1.52 & 0.15 & 451 &  90 \\
 23.25 & 1.69 & 0.21 & 522 &  87 \\
 23.75 & 1.83 & 0.22 & 569 &  84 \\
 24.25 & 1.90 & 0.21 & 636 &  80 \\
 24.75 & 2.05 & 0.36 & 586 &  74 \\
 25.25 & 2.16 & 0.23 & 532 &  54 \\ \hline
\end{tabular}
\vspace{0.5cm}
\label{tab2}
\end{table}

\begin{table}
\centering
\caption{Main sequence fiducial points, luminosity function and
completeness for NGC\,6352}
\begin{tabular}{cccccc} \hline
{\em $m_{814}$} & {\em $m_{606}-m_{814}$} & {\em $\sigma$} &{\em
$N_{o}$} & {\em Compl. (\%)}\\
\hline
 17.25 & 0.77 & 0.03 & 35 & 100 \\
 17.75 & 0.74 & 0.02 & 67 &  99 \\
 18.25 & 0.75 & 0.03 & 95 &  98 \\
 18.75 & 0.78 & 0.04 & 119 &  97 \\
 19.25 & 0.83 & 0.07 & 172 &  96 \\
 19.75 & 0.90 & 0.07 & 250 &  95 \\
 20.25 & 1.00 & 0.11 & 267 &  93 \\
 20.75 & 1.17 & 0.14 & 297 &  94 \\
 21.25 & 1.33 & 0.15 & 294 &  93 \\
 21.75 & 1.53 & 0.27 & 368 &  90 \\
 22.25 & 1.68 & 0.27 & 447 &  87 \\
 22.75 & 1.83 & 0.26 & 527 &  82 \\
 23.25 & 1.87 & 0.27 & 464 &  78 \\
 23.75 & 1.94 & 0.29 & 379 &  73 \\
 24.25 & 1.88 & 0.42 & 353 &  60 \\ \hline
\end{tabular}
\vspace{0.5cm}
\label{tab3}
\end{table}

\begin{figure*}
\centering
\includegraphics[width=16cm]{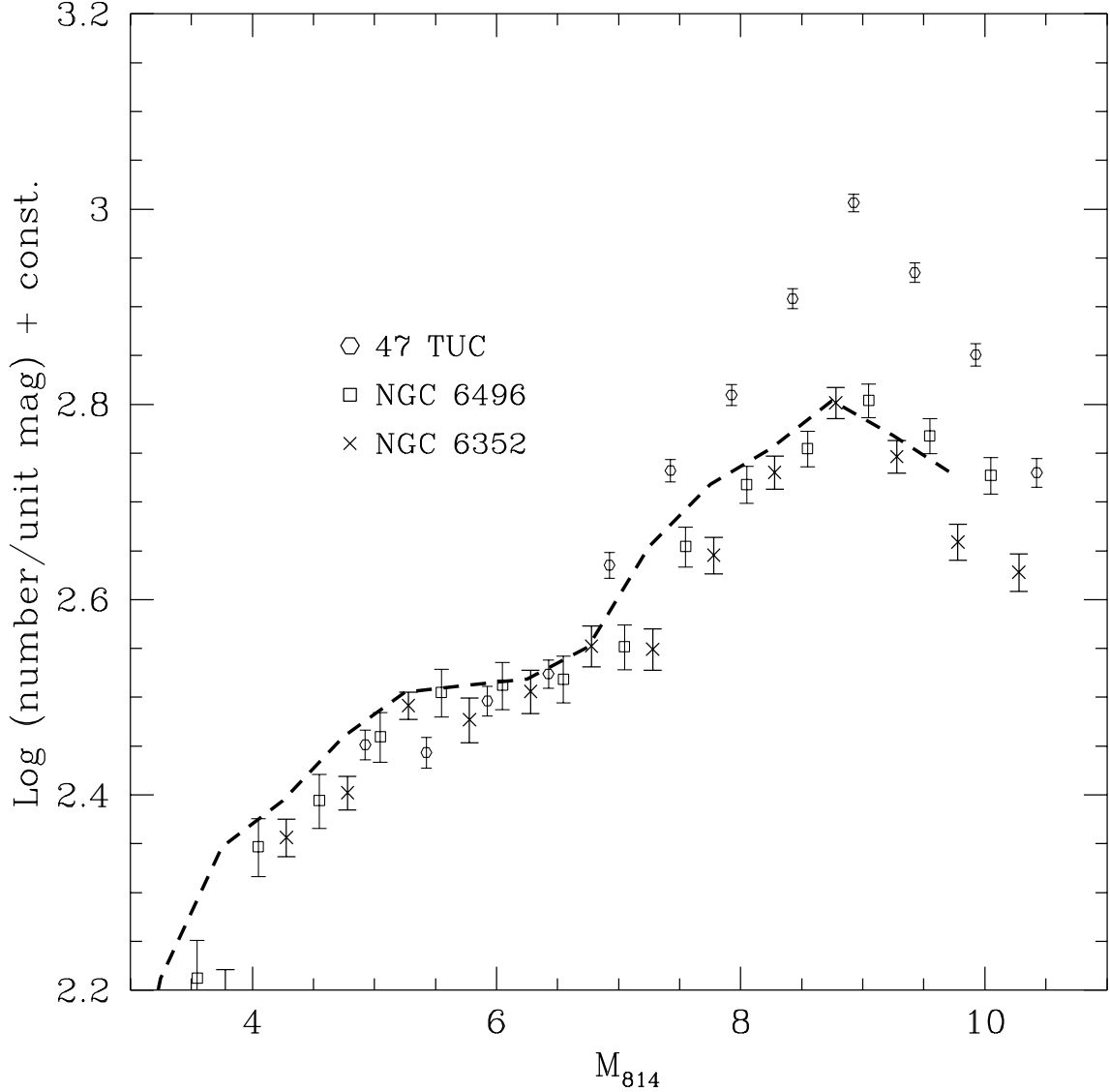}
\caption{Measured LF of NGC\,6496 (boxes) and NGC\,6352
(crosses) as compared with that of NGC\,104 (circles). The dashed line
shows the LF of NGC\,6496 obtained by adopting the distance modulus
$(m-M)_I=15.5$ as derived from the catalogue of Harris (1996)}
\label{fig6}
\end{figure*} 

We have divided the MS into magnitude bins, each spanning $0.5$\,mag
from $m_{814} = 18$ to $m_{814} = 25.5$, and from $m_{814} = 17$ to
$m_{814} = 24.5$ for NGC\,6496 and NGC\,6352, respectively. By adopting
a sigma clipping criterion (De Marchi \& Paresce 1995b), from the CMD
of Figure\,\ref{fig3} we have measured the LF by counting the
objects in $0.5$ mag bins and within $\pm 2.5$ times the
$m_{606}-m_{814}$ colour standard deviation ($\sigma$) around the MS
ridge line.  As regards the decontamination of the LF from the field
stars, we have assumed that the field objects fill uniformly the colour
range $0.3 < m_{606}-m_{814} < 3$. For each magnitude bin, we have
estimated the number of stars with colour in this range, but outside
the $5\sigma$ wide MS belt. By dividing the measured value by the
corresponding colour range and multiplying the result by the MS colour
width, we have obtained the number of field stars inside the MS. The LF
corrected for incompleteness and field star contamination is shown in
Figure\,\ref{fig6}, together with that of 47\,Tuc, as a function of the
absolute $M_{814}$ magnitude. The LF of NGC\,6496 and NGC\,6352 rise
continuously, all the way up to $M_{814}\simeq9$, where they reach
their maximum, and drop to the detection limit at $M_{814} \simeq 10$.
Taken at face value, after its peak the LF of NGC\,6352 seems to
decrease somewhat more steeply than that of NGC\,6496, as indicated by
the $1 \sigma$ error bars. We should note, however, that the difference
between the two LF in practice vanishes if we consider the uncertainty
brought about by our $0.5$\,mag wide bins along the abscissa. Moreover,
if we had assumed for the distance modulus of NGC\,6496 the canonical
value of $(m-M)_I=15.5$ as given by Harris (1996), the LF of the two
clusters would have been in reasonable agreement with one another (see
dashed line in Figure\,\ref{fig6}). The value of $(m-M)_I=15.2$ that we
have used was suggested by the remarkably good match between the MS
fiducial lines of the two clusters that it would provide in the CMD
(see Section\,4). The value $(m-M)_I=15.5$ would imply that
NGC\,6496 should be slightly more metal rich than NGC\,6352, a
hypothesis that we cannot certainly rule out at the present level of
accuracy. We, therefore, conclude that the LF of NGC\,6496 and
NGC\,6352 are substantially the same.

To best fit the LF of the three clusters under examination, we have to
fold an input MF through the derivative of the M--L relation appropriate
for the cluster metallicity. We adopted the M--L relation obtained from
the $10$\,Gyr isochrones of Baraffe et al. (1998) with [M/H]$=-0.5$ and
$Y=0.25$ in the F814W filter for the reasons outlined above. When
looking at the admittedly rather poor fit of the theoretical isochrones
to the MS loci of these clusters in Figure\,\ref{fig3} below
$M_{814}\simeq 7$, it is fair to wonder whether the resulting M--L
relations are reliable at all. We have, thus, examined how an
improvement in the treatment of the TiO opacity could alter the M--L
relation and give a better fit. For the solar mixture, Baraffe et al.
(1998) provide isochrones with and without TiO opacity enhancement. We
obtained M--L relations with and without this TiO effect and realised
that the resulting MF were substantially the same, thus suggesting
(although not proving) that in the case of [M/H]$=-0.5$ as well the
adoption of the uncorrected M--L should give reliable results. Moreover,
the quoted numerical experiments show that the effects of the
uncertainty of the TiO opacity values on the M--L relation should be
negligible when compared to the smoothing of the observed LF caused by
the $0.5$ magnitude binning.

\begin{figure}
\resizebox{\hsize}{!}{\includegraphics{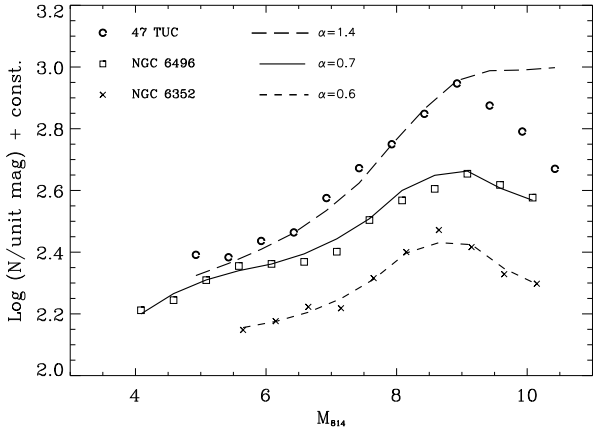}}
\caption{Observed LF of NGC\,6496
(squares), NGC\,6352 (crosses), and 47\,Tuc (open circles) compared with 
those obtained by folding a power-law MF of appropriate index through the
M--L relation of Baraffe et al. (1998) for metallicity [M/H]$=-0.5$}
\label{fig7}
\end{figure}

We compare in Figure\,\ref{fig7} the observed LF with that obtained
when a model MF is folded through the M--L relation. The squares and
crosses refer, respectively, to NGC\,6496 and NGC\,6352. We have made
here the simplifying assumption that the MF of the stellar population
near the cluster half-mass radius is represented by a power-law, i.e.
that the mass distribution function takes on the form $dN/dm \propto
m^{-\alpha}$, finding that values of $\alpha=0.7\pm0.1$ and $0.6\pm0.1$
are adequate for NGC\,6496 and NGC\,6352, respectively. As a
short-dashed line in Figure\,\ref{fig7} we also show the model LF that
best fits the observations of 47\,Tuc, the prototypical cluster of
intermediate metallicity. It is well known that a simple power-law MF
cannot adequately reproduce the MS LF of 47\,Tuc over the whole
observed range of luminosity and that a log-normal distribution in mass
(Paresce \& De Marchi 2000) or a tapered power-law function (De Marchi
et al. 2003) give a better fit over the whole mass range.
Nevertheless, for the sake of the discussion that follows, it is
sufficient to limit the comparison amongst the three MF within the
range $0.3 < m/m_\odot < 0.8$, where a single exponent power-law is the
simplest and, hence, the preferred assumption. Figure\,\ref{fig7}
leaves no doubt that the MF of 47\,Tuc, with $\alpha=1.4$, is twice as
steep as that of the other two clusters.
 
\section{Discussion and conclusions}

To the expert eye, this discrepancy would probably not come as a
surprise, since it is the natural consequence of the already different
LF seen in Figure\,\ref{fig6}. To be precise, however, the difference
in the LF per se would not necessarily imply that also the underlying
MF should be different since 47\,Tuc has a lower metal content than
the other two objects. Thus, stars of equal mass would appear slightly
brighter in 47\,Tuc due to the different M--L relation, therefore
changing the shape of the observed LF. In fact, it is well understood
that the most notable feature in the LF of a cluster, i.e. the
brightness at which it peaks, shifts to fainter magnitudes with
increasing metallicity (see e.g. D'Antona 1998; Kroupa \& Tout 1997;
von Hippel et al. 1996; De Marchi \& Paresce 1995b) to the point that,
if the distance to a cluster were known, the magnitude of the peak
could be used as a metallicity indicator.

The application of the M--L relations appropriate to each cluster has,
however, removed the signature of the metallicity. As explained in
Section\,5, although some uncertainties exist as to the exact shape of
the M--L relationship in the range $-0.5 < {\rm [Fe/H]} < 0.0$ (Baraffe
et al. 1998), they cannot be held responsible for the large discrepancy
in the MF seen in Figure\,\ref{fig3}. We must, therefore, accept that
NGC\,6352 and NGC\,6496 have MF which are inherently different from
that of 47\,Tuc and from the dozen lower metallicity halo clusters
studied to date in detail with the HST (Paresce \& De Marchi 2000).

The effects of energy equipartition and the resulting mass segregation
have been observed and studied in detail in several clusters and can be
modelled satisfactorily (47\,Tuc, Paresce et al. 1994; NGC\,7078, De
Marchi \& Paresce 1996; NGC\,6397, King et al. 1995, De Marchi et al.
2000; NGC\,6121, Pulone et al. 1999; NGC\,6656, Albrow et al. 2001). It
is today understood that, near the cluster's half-mass radius, the
local MF faithfully reflects the global MF of the whole cluster (as
originally suggested by Richer et al. 1991). Since the observations of
NGC\,6352 and NGC\,6496 on which we here report were conducted at their
half-mass radius (see Section\,2), the effects of internal dynamics can
be safely excluded from the picture.

On the other hand, the action of the Galactic tidal field can be
equally strong in reshaping the MF of a cluster through disc and bulge
shocks during its repeated passages at the perigalacticon. For example,
the MF of NGC\,6712 measured near the half-mass radius (De Marchi et
al. 1999; Andreuzzi et al.  2001) bears no resemblance to that of any
other cluster studied so far, in that if falls steeply with decreasing
mass. Pal\,5 (Grillmair \& Smith 2001) and E\,3 (van den Bergh et al.
1980) might have undergone a similar, although less severe fate.  It
is, thus, fair to wonder whether the MF of NGC\,6352 and NGC\,6496 were
similarly affected by the Galaxy.

Gnedin \& Ostriker (1997) and Dinescu et al. (1998) used the mass, size
and space motion parameters of most Galactic globular clusters to infer
the magnitude of their past dynamical interaction with the potential
field of the Galaxy. They define a useful parameter, the time to
disruption ($t_d$), which should directly correlate with the strength
and extent of the overall tidal interaction and, thus, ultimately with
the shape of the present day MF. However, whilst it is true that for
NGC\,6712 they predict a probability of disruption about two orders of
magnitude higher than average, Paresce \& De Marchi (2000) have shown
that no correlation exists between $t_d$ and the shape of the MF of the
12 clusters that they studied in detail with the HST. The value of
$t_d$ must thus be used with care and we shall do so here to show, with
a very simple reasoning, that neither NGC\,6352 nor NGC\,6496 can have
been severely affected by tidal stripping. If the theory of Gnedin \&
Ostriker (1997) is indeed correct, the destruction rate for NGC\,6496
is about 300 times larger than that of NGC\,6352, yet the two clusters
have the same global MF! So, if we want these clusters to have had an
IMF originally as steep as that of 47\,Tuc, we cannot hold tidal
stripping accountable for the alleged flattening. Moreover, postulating
that the two IMF were originally different from one another but that,
through dynamical evolution, they have been brought into agreement
would look highly contrived, in light of the widely different dynamical
histories of the two clusters. Lacking any other viable explanation, it
is more logical to conclude that the IMF of these clusters did not
change with time, i.e.  that it was already flatter at birth and the
same for both.

This conclusion, however, does still not explain why the IMF of
NGC\,6352 and NGC\,6496 should be flatter than that of the less metal
rich GC unless, of course, it is the metallicity itself that governs
the shape of the IMF. Claims of an apparent correlation between the
shape of the MF and cluster metallicity have been put forth in the past
by McClure et al. (1986), who analysed a set of nine, relatively deep
GC LF obtained from the ground and noticed a shallower slope for richer
clusters, in the same sense that we report here. More recent data,
however, do not confirm these results (Paresce \& De Marchi 2000). Von
Hippel et al. (1996) went even further and studied how metallicity
affects the shape of the IMF of clusters of various ages (from young
open cluster to globulars), concluding in favour of an invariant IMF.

Whether stars form through the hierarchical fragmentation of a
proto-stellar cloud (Elmegreen 1999) or through the accretion of smaller
clumps (Bonnell et al. 2001), once the pre-MS phase begins, the
metallicity and ensuing atmospheric opacity will eventually determine
the limiting mass for stable Hydrogen burning (see e.g. D'Antona \&
Mazzitelli 1996) and, thus, the value of the IMF peak mass. In other
words, the metal content of the natal cloud will affect its temperature
and pressure, on which the characteristic mass eventually depends
(Larson 1998). If turbulent fragmentation prevails below $\sim
1$\,M$_\odot$, its dependence on the cooling rate and, hence, on the
metallicity could affect the shape of the IMF, but no clear prediction
as to the results of this process exists (Larson 2002). Moreover, the
observational evidence seems overwhelming that the IMF is universal and
independent of the total mass, density, age and metallicity of the
stellar population (Scalo 1998; Gilmore 2001; but see also the
cautionary remark in Kroupa 2001).

NGC\,6352 and NGC\,6496 would seem, at least apparently, to violate
this rule. We have, however, so far ignored the important contribution
of binaries (Kroupa 2001). Other examples exist of relatively old and
metal rich systems that display a rather flat MF: the bulge (Holtzman
et al. 1998; Zoccali et al. 2000), the spheroid (Gould et al. 1997) as
well as the Galactic disc in the solar neighbourhood (Kroupa et al.
1993; Ried \& Gizis 1997; Gould et al. 1997). In all cases, the MF of
these systems, as derived directly from the observed LF, is rather flat
if no account is taken of the presence of binaries: for instance, for
stars less massive than $\sim 0.5$\,M$_\odot$ Gould et al. (1997) find
$\alpha = 0.56$ in the disc and $\alpha = 0.75$ in the spheroid.
Ignoring the binaries, however, results in the underestimate of the
real IMF slope by an amount that depends on the slope itself and on the
binary fraction (Sagar \& Richtler 1991; Kroupa 2001) and which can be
of order $\Delta\alpha \simeq 0.5$ or higher. As Holtzman et al. (1998)
show, a measured MF slope of $\alpha\simeq 0.7$ can be in reality as
steep as $\alpha\simeq 1.3$ if the binary fraction is of order 50\,\%.

The age and metallicity of the stellar population in NGC\,6496 and
NGC\,6352 should be rather similar to those of the stars in the bulge,
as one can judge for instance from the similarity of their CMD and
particularly of their curved red giant branch and clumped horizontal
branch (Richtler et al. 1994; Fullton et al. 1995; Zoccali et al.
2000). It is, then, reasonable to apply to their MF a correction
similar to that derived by Holtzman et al. (1998), which would in turn
bring them into good agreement with that of 47\,Tuc and of the other,
lower metallicity halo clusters (Paresce \& De Marchi 2000).

At this juncture, one might wonder whether a similar correction for
binaries should be applied to all GC, including 47\,Tuc. Albrow et al.
(2001), however, have recently conducted high precision differential
time series photometry on $\sim 46,000$ stars in the core of this
cluster and have concluded that the binary fraction is on average $\sim
13\,\%$ within $\sim 4$ core radii ($r_c$), with indications that it
drops to $\sim 8\,\%$ or less outside of $2.5\,r_c$. A correction to the
IMF in this case is, therefore, not necessary. The only notable
signature of binaries on the observed CMD is a vertical broadening of
the MS (of up to $0.75$\,mag along the magnitude axis for equal mass
systems), which should be best observed where the MS slope is not too
steep in the CMD, or where it changes abruptly near the MS kinks due to
changes in the main source of opacity (see, e.g., the CMD of NGC\,6752
by Rubenstein \& Bailyn 1997). With the advent of high precision HST
photometry, this test has become possible (De Marchi \& Paresce 1995b;
Elson et al. 1995; Cool et al. 1996) and has revealed the paucity of
these objects. Yet, in only a few cases, for example NGC\,6752
(Rubenstein \& Bailyn 1997), has the photometry been sufficiently
precise to detect this binary sequence.  

The MS of NGC\,6496 and NGC\,6352 is, however, much steeper than that
of NGC\,6752 in the magnitude range where the photometric uncertainty
would allow us to see the effects of a broadening. Below $m_{606}\simeq
23$ the MS of NGC\,6496 flattens out, but here the photometric
uncertainty dominates (see Table\,2). Furthermore, since the vertical
displacement of a binary with respect to the MS in the CMD depends on
the mass ratio $q=m_2/m_1$, it will be easier to detect nearly equal
mass systems, which are however believed to be a small fraction of the
total, at least according to Duquennoy \& Mayor (1991). These authors
investigate the multiplicity of nearby G dwarfs and find that the
secondary mass distribution peaks at around $q\simeq 0.25$, where the
solar neighbourhood's IMF has its maximum (Kroupa et al. 1993), thereby
suggesting that binaries should form some time after the formation of
single stars.  If the same mass distribution applies to the binaries in
NGC\,6496 and NGC\,6352, their signature on the MS will remain
undetected, at least at the present level of photometric accuracy,
since a typical $\sim 0.25$\,M$_\odot$ companion would add
insignificantly to the brightness of a more massive primary. Although
for late M dwarf primaries the $q$ distribution seems somewhat flatter
than for G stars (Fischer \& Marcy 1992), again in agreement with a
binary formation scenario through random sampling from the same IMF,
our photometric uncertainty, the contamination due to field objects and
the steepening of the MS in the CMD plane prevent us from reliably
probing this region of the parameter space.
  
Whilst NGC\,6352 and NGC\,6496 are the only two clusters in the bulge
for which a deep MF has been measured so far, at this stage one could
tentatively conclude that the main effect of the metallicity difference
between these and the halo clusters may indeed be the binary
frequency. The proposed universality of the IMF would not seem to be
affected by this conclusion, but it must not be disconnected from the
process of binary formation. In order for the final stellar IMF to be
the same regardless of the metal content, it would require that stars
are all born single and that only later they aggregate in multiple
systems within the natal cloud (where they are observed), in a way
which however depends on the metallicity. That the properties of
binaries can be better explained through a ``two step'' IMF has been
discussed recently by Durisen et al. (2001). Thus, it cannot be
excluded, for instance, that the stronger UV flux and more powerful
winds of OB stars in metal poor clusters would not only hinder the
formation of low mass objects (Larson 1998), but also discourage their
interaction and aggregation into multiple systems. This scenario agrees
with the finding (Duquennoy \& Mayor 1991; Fischer \& Marcy 1992) that,
in the solar neighbourhood, binary formation seems to happen through
random association of two stars drawn from the same IMF. An
alternative proposal is also possible in which, after the initial
collapse, clumps form which further fragment into single stars or
multiple systems depending on the cloud's metallicity. This hypothesis
would, however, be less favoured because in order to conserve the
uniformity of the stellar IMF with metallicity the fragmentation
mechanism should adapt itself to the metal content in a way which is
presently not understood.

In spite of the rather speculative nature of the previous paragraph, we
stress here that our result is the first tentative indication of a
large fraction of binaries in some globular clusters.

\section{Summary}
 
The main results of this paper can be summarised as follows:

\begin{enumerate}

\item Two regions located at the respective half-light radius of
NGC\,6496 and NGC\,6352 were observed with the WFPC2 on board the HST.
The photometry of the WF chips, carried out in the F814W and F606W
bands, extends down to $m_{814} \sim 25.5$ corresponding to $M_{814}
\simeq 10.5$ with completeness in the raw data always above $\sim
50\,\%$.

\item The MS fiducial lines of NGC\,6496 and NGC\,6352 show a similar
shape within the measurement errors, thus strongly suggesting a
similar metal content, which appears at least $0.2-0.3$ dex higher
in [M/H] than that of 47\,Tuc. Accordingly, neither object should 
be considered a member of the halo Galactic globular cluster system.

\item No theoretical M--L relationships are available that reproduce the
shape of the MS fiducial line over the whole range of luminosity
spanned by the observations. According to Baraffe et al. (1998), the
offset observed at low luminosities (and masses) is a shortcoming
reflecting the inadequate treatment of the TiO molecule, which becomes
important for metallicity higher than [M/H]$\simeq -0.5$. This effect
is less severe in the reddest bands.

\item The I-band LF of NGC\,6496 and NGC\,6352, obtained from the CMD
using both the colour and magnitude information and corrected for
incompleteness were compared to the LF of 47\,Tuc. Although
all three LF feature a peak at a similar magnitude level ($M_{814}
\simeq 9$), that of 47\,Tuc rises more steeply to the maximum.

\item Using the M--L of Baraffe et al. (1998), the power-law MF that
best fit the observations in the range $0.3 < m/m_\odot < 0.8$ are
derived and compared, finding an index $\alpha=0.7$ and $0.6$ for
NGC\,6496 and NGC\,6352, respectively. The MF of 47\,Tuc is twice as
steep over the same mass range, with $\alpha=1.4$ and similarly steep is
the MF of other halo GC.

\item We investigate the origin of this discrepancy and conclude that a
dynamical origin for it is highly unlikely if present theories are
correct. We suggest instead that it could be the result of a large
binary fraction amongst the stars of NGC\,6496 and NGC\,6352. A
binary fraction of $\sim 50\,\%$ would produce a stellar MF in
agreement with that of 47\,Tuc and of the low metallicity halo
clusters. Because of their location and chemical properties, both
NGC\,6496 and NGC\,6352 are likely members of the bulge, where such a
binary fraction is accepted (Holtzman et al. 1998). If this result is
confirmed, it would imply that the metallicity affects the way in which
binaries form and aggregate, but not the fragmentation process of the
natal cloud.  A larger sample of intermediate and high metallicity GC
is needed to address these issues in more detail.

\end{enumerate}

\begin{acknowledgements}

It is our pleasure to thank I. Baraffe for useful discussions. L.
Pulone is grateful to ESO, where he conducted part of this work, for
the hospitality of their visitor programme. We are very thankful to an
anonymous referee whose comments have considerably improved the
presentation of this paper. This work has been partly supported by
MURST/COFIN 2000 under the project: ``Stellar observables of
cosmological relevance.''

\end{acknowledgements}

\end{document}